\documentclass[runningheads]{llncs}
\usepackage{booktabs} 
\usepackage{tabularx}
\newcommand{\mpara}[1]{\medskip\noindent{\bf #1}}
\usepackage[misc]{ifsym}
\usepackage{url} 
\usepackage{xcolor}

\usepackage{graphicx}
\usepackage{multirow}
\usepackage{booktabs}

\usepackage{colortbl}
\usepackage{pifont}
\newcommand{\cmark}{\ding{51}}
\newcommand{\xmark}{\ding{55}}

\PassOptionsToPackage{table}{xcolor} %
\usepackage[table]{xcolor}
\usepackage{hyphenat}
\usepackage{url}
\usepackage{xspace} %

\usepackage{amsthm}
\usepackage{amsmath}
\usepackage{mathtools}
\usepackage{amsfonts}
\usepackage{algorithm}
\usepackage{algorithmicx}
\usepackage[noend]{algpseudocode}
\usepackage{chngcntr} %
\usepackage{todonotes}

\usepackage[font=small]{caption}

\usepackage{balance}
\usepackage{enumitem}
\usepackage{soul}
\usepackage{tikz}
\usepackage{tabularx}
\usetikzlibrary{fit,positioning,calc,shapes}
\usepackage{pgfplots}
\pgfplotsset{compat=1.13}
\usepgfplotslibrary{colorbrewer}
\usepackage{multirow}
\usepackage[caption=false]{subfig}
\usepackage{breqn}
\usepackage{caption}
\usepackage{pifont}
\usepackage{stfloats}
\usepackage{booktabs}

\definecolor{cycle1}{RGB}{228, 26, 28}
\definecolor{cycle2}{RGB}{55, 126, 184}
\definecolor{cycle3}{RGB}{77, 175, 74}
\definecolor{cycle4}{RGB}{152, 78, 163}
\definecolor{cycle5}{RGB}{255, 127, 0}
\definecolor{cycle6}{RGB}{153, 153, 153}%
\definecolor{cycle7}{RGB}{166, 86, 40}
\definecolor{cycle8}{RGB}{247, 129, 191}

\newcommand{\mehmark}{\mbox{\cmark\textsubscript{\kern-0.45em\tiny\xmark}}}

\newcommand{\deepwalk}{\textsf{DeepWalk}\xspace}

\newcommand{\lineemb}{\textsf{LINE}\xspace}
\newcommand{\verseemb}{\textsf{VERSE}\xspace}

\newcommand{\nodetovec}{\textsf{Node2vec}\xspace}

\newcommand{\hoppe}{\textsf{HOPE}\xspace}
\newcommand{\hope}{\textsf{HOPE}\xspace}
\newcommand{\nerd}{\textsf{NERD}\xspace}

\newcommand{\app}{\textsf{APP}\xspace}
\newcommand{\edgednn}{\textsf{EdgeDNN}\xspace}

\newcommand{\fromto}{\longrightarrow}
\renewcommand{\to}{\fromto}

\DeclarePairedDelimiterX{\norm}[1]{\lVert}{\rVert}{#1}
\theoremstyle{definition}
\newtheorem{defn}{Definition}

\newcommand{\pubmed}{\textbf{\textsf{PubMed}}}
\newcommand{\cora}{\textbf{\textsf{Cora}}}
\newcommand{\epi}{\textbf{\textsf{Epinion}}}
\newcommand{\twitter}{{\textbf{\textsf{Twitter}}}}
\newcommand{\cocit}{{\textbf{\textsf{CoCit}}}}

\newcolumntype{H}{>{\setbox0=\hbox\bgroup}c<{\egroup}@{}}

\begin{document}
\title{Node Representation Learning for Directed Graphs}
\toctitle{Node Representation Learning for Directed Graphs}
%
%
\author{Megha Khosla~{\Letter} \and
Jurek Leonhardt \and
Wolfgang Nejdl\and
Avishek Anand
}
\authorrunning{M. Khosla et al.} 
\tocauthor{Megha Khosla,
Jurek Leonhardt,
Wolfgang Nejdl,
Avishek Anand}

%
\institute{L3S Resaerch Center, Hannover, Germany
\email{\{khosla,leonhardt,nejdl,anand\}@l3s.de}}
\maketitle              
\begin{abstract}
We propose a novel approach for learning node representations in directed graphs, which maintains separate views or embedding spaces for the two distinct node roles induced by the directionality of the edges.
We argue that the previous approaches either fail to encode the edge directionality or their encodings cannot be generalized across tasks.
With our simple \emph{alternating random walk} strategy, we generate role specific vertex neighborhoods and train node embeddings in their corresponding source/target roles while fully exploiting the semantics of directed graphs. 
We also unearth the limitations of evaluations on directed graphs in previous works and propose a clear strategy for evaluating link prediction and graph reconstruction in directed graphs. 
We conduct extensive experiments to showcase our effectiveness on several real-world datasets on link prediction, node classification and graph reconstruction tasks. We show that the embeddings from our approach are indeed robust, generalizable and well performing across multiple kinds of tasks and graphs. We show that we consistently outperform all baselines for node classification task. 
In addition to providing a theoretical interpretation of our method we also show that we are considerably more robust than the other directed graph approaches.

\end{abstract}
%
%
%
%
%
%
\section{Introduction}

Unsupervised representation learning of nodes in a graph refers to dimensionality reduction techniques where nodes are embedded in a continuous space and have dense representations.
Such node embeddings have proven valuable as representations and features for a wide variety of prediction and social network analysis tasks such as link prediction \cite{liben2007link}, recommendations \cite{ying2018graph}, vertex label assignment, graph generation \cite{drobyshevskiy2017learning} etc.

However most of the recent node embedding methods have been focused on undirected graphs with limited attention to the directed setting.
Often valuable knowledge is encoded in directed graph representations of real-world phenomena where an edge not only suggests relationships between entities, but the directionality is often representative of important asymmetric semantic information. 
Prime examples are follower networks, interaction networks, web graphs, and citation networks among others. 


Most of the approaches in this regime~\cite{tang2015line,Grover:2016,Perozzi:2014} focus on the goal of preserving neighborhood structure of nodes when embedding one space into another, but suffer from some key limitations when representing directed graphs.
First, most of these node embedding techniques operate on a single embedding space and distances in this space are considered to be symmetric. 
Consequently, even though some of the approaches claim to be applicable for directed graphs, they do not respect the \textbf{asymmetric roles} of the vertices in the directed graph. For example, in predicting links in an incomplete web graph or an evolving social network graph, it is more likely that a directed link exists from a less popular node, say \texttt{Max Smith}, to a more popular node, say an authoritative node \texttt{Elon Musk}, than the other way around. Algorithms employing single representations for nodes might be able to predict a link between \texttt{Elon Musk} and \texttt{Max Smith} but cannot predict the direction. 

Secondly, approaches like \app~\cite{AAAI1714696} overcome the first limitation by using two embedding spaces but are unable to differentiate between directed neighborhoods where these neighborhoods can be distinguished based on reachability. For example, for a given node $v$ there exists a neighborhood which is reachable from $v$ and there exists another type of neighborhood to which $v$ is reachable.
More acutely, many nodes with zero outdegree and low indegree might not be sampled because of the training instance generation strategy from its  random walk following only outgoing edges. 
This renders such approaches \textbf{not to be robust}, a desirable and important property for unsupervised representations, for several real-world graphs.

Finally, works like \hoppe~\cite{ou2016asymmetric} rely on stricter definitions of neighborhoods dictated by proximity measures like Katz~\cite{katz1953new}, Rooted PageRank etc. and cannot be generalized to a variety of tasks. 
In addition, they do not scale to very large graphs due to their reliance on matrix decomposition techniques. 
Moreover, the accuracy guarantees of \hoppe rely on low rank assumption of the input data. 
Though not completely untrue for real world data, \emph{singular value decomposition}(SVD) operations used in matrix factorization methods are known to be sensitive even for the case of a single outlier \cite{ammann1993robust}. 
We later empirically demonstrate in our experiments that \hoppe can not be easily adapted for the node classification task as it is linked to a particular proximity matrix.

We argue that the utility and strength of unsupervised node representations is in their (1) robustness across graphs and (2) flexibility and generalization to multiple tasks and propose a simple yet robust model for learning node representations in directed graphs.


\mpara{Our Contribution.}
We propose a robust and generalizable approach for learning \textbf{N}ode \textbf{E}mbeddings \textbf{R}especting \textbf{D}irectionality (\nerd) for directed and (un)--weighted graphs. \nerd aims at learning representations that maximize the likelihood of preserving node neighborhoods. But unlike the previous methods, it identifies the existence of \textbf{two} different types of node neighborhoods; one in its source role and the other in its target role. We propose an \emph{alternating random walk} strategy to sample such node neighborhoods while preserving their respective role information. Our alternating walk strategy is inspired from SALSA~\cite{lempel2001salsa} which is a  stochastic variation of the HITS~\cite{Kleinberg:1999} algorithm and also identifies two types of important nodes in a directed network: {\em hubs} and {\em authorities}. Roughly speaking, the paths generated with our alternating random walks alternate between hubs (source nodes) and authorities (target nodes), thereby sampling both neighboring hubs and authorities with respect to an input node. 
From a theoretical perspective we derive an equivalence for \nerd's optimization in a matrix factorization framework. In addition, we also unearth the limitations of earlier works in the evaluation of models on directed graphs and propose new evaluation strategies for Link Prediction and Graph Reconstruction tasks in directed graphs. Finally we perform exhaustive experimental evaluation that validates the robustness  and generalizability of our method.

\section{Related Work}
\label{sec:related-work}
Traditionally, undirected graphs have been the main use case for graph embedding methods. Manifold learning techniques \cite{NIPS2001_1961}, for instance, embed nodes of the graph while preserving the local affinity reflected by the edges. Chen et al.~\cite{chen2007directed} explore the directed links of the graph using random walks, and propose an embedding while preserving the local affinity defined by directed edges. Perrault-Joncas et al.~\cite{perrault2011directed} and Mousazadeh et al.~\cite{Mousazadeh:2015} learn the embedding vectors based on Laplacian type operators and preserve the asymmetry property of edges in a vector field.
 
Advances in language modeling and unsupervised feature learning in text inspired their adaptations~\cite{Grover:2016,Perozzi:2014,cao2015grarep,tang2015line,tsitsulin2018verse} to learn node embeddings where the main idea is to relate nodes which can reach other similar nodes via random walks.  \deepwalk~\cite{Perozzi:2014}, for instance, samples truncated random walks from the graph, thus treating \emph{walks} as equivalent of sentences, and then samples node-context pairs from a sliding window to train a \emph{Skip-Gram} model \cite{mikolov2013distributed}. 
 \nodetovec \cite{Grover:2016}, on the other hand, uses a \emph{biased} random walk procedure to explore diverse neighborhoods. \lineemb \cite{tang2015line} which preserves first and second order proximities among the node representations can also be interpreted as embedding nodes closer appearing in random walk of length 1.
 \verseemb \cite{tsitsulin2018verse} learns a single embedding matrix while encoding similarities between vertices sampled as first and last vertex in a PageRank style random walk. 
 
Other works \cite{Cao:2016,wang2016structural,kipf2016variational,arga} investigate deep learning approaches for learning node representations. 
Like most of the other methods, they also use a single representation for a node, hence ignoring the asymmetric node roles. Other downsides of these deep learning approaches are the computationally expensive optimization and elaborate parameter tuning resulting in very complex models. 


\mpara{Asymmetry preserving approaches.} To the best of our knowledge, there are only two works~\cite{ou2016asymmetric,AAAI1714696} which learn and use two embedding spaces for nodes, one representing its embedding in the source role and the other in the target role. 
Note that \cite{learnEdge2017} does not preserve asymmetry for the nodes, which is the main theme of this work (more comparisons and discussions on this method can be found in the Appendix). 
\hoppe~\cite{ou2016asymmetric} preserves the asymmetric role information of the nodes by approximating high-order proximity measures like Katz measure, Rooted PageRank etc. 
Basically they propose to decompose the similarity matrices given by these measures and use the two decompositions as representations of the nodes. \hoppe cannot be easily generalized as it is tied to a particular measure. 
\app~\cite{AAAI1714696} proposes a random walk based method to encode rooted PageRank proximity. 
Specifically, it uses directed random walks with restarts to generate training pairs. Unlike other \deepwalk style random walk based methods, \app does not discard the learnt context matrix, on the other hand it uses it as a second (target) representation of the node. 
However, the random walk employed sometimes is unable to capture the global structure of the graph. Consider a directed graph with a prominent hub and authority structure where many authority nodes have no outgoing links. In such a case any directed random walk from a source node will halt after a few number of steps, irrespective of the stopping criteria. 
In contrast our alternating random walks also effectively sample low out-degree vertices in their target roles, thereby exploiting the complete topological information of a directed graph.




\section{The NERD Model}
\label{sec:model}
\vspace{-2mm}
Given a directed graph $G=(V,E)$ we aim to learn $d$-dimensional ($d << |V|$) representations, $\Phi_s$ and $\Phi_t$ , such that the similarities between vertices in their respective source and target roles are preserved.
We argue that that two vertices can be similar to each other in three ways: (i) both nodes in source roles (both pointing to similar authorities) (ii) both the nodes in target roles (for example both are neighbors of similar hubs) (iii) nodes in source-target roles (hub pointing to a authority). We extract such similar nodes via our \emph{alternating random walk} strategy which alternates between vertices in opposite roles. For every vertex two embedding vectors are learnt via a single layer neural model encoding its similarities to other vertices in source and target roles. Alternatively, \nerd can be interpreted as optimizing first order proximities in three types of \emph{computational graphs} it extracts from the original graph via alternating random walks. We elaborate on this alternate view while explaining our learning framework in Section~\ref{sec:learning}.
 
 \mpara{Notations.} We first introduce the notations that would also be followed in the rest of the paper unless stated otherwise. Let $G=(V,\vec{E})$ be a directed weighted graph with $N$ nodes and $M$ edges. Let $w(e)$ denote the weight of edge $e$ and $vol(G)=\sum_e w(e)$. For any vertex $v\in V$ let $d^{out}(v)$ denote the total outdegree of $v$, i.e.\ the sum of weights of the outgoing edges from $v$.  Similarly $d^{in}(v)$ denotes the total indegree of $v$. For unweighted graphs, we assume the weight of each edge to be $1$. Let $\Phi_s(v)$ and $\Phi_t(v)$ represent the respective embedding vectors for any node $v\in V$ in its role as source and target respectively. Let $P^{in}$ and $P^{out}$ denote the input and output degree distributions of $G$ respectively. Specifically $P^{in}(v)= d^{in}(v)/ vol(G)$ and $P^{out}(v)= d^{out}(v)/vol(G)$. We remark that the terms \emph{vertex} and \emph{node} are used interchangeably in this work.

\subsection{Alternating Walks}
We propose two alternating walks which alternate between source and target vertices and are referred to as \emph{source} and \emph{target} walks respectively. To understand the intuition behind these walks, consider a directed graph $G=(V,\vec{E})$ with $N$ nodes. Now construct a copy of each of these $N$ nodes and call this set $V_c$. Construct an undirected bipartite graph $G'=(V \cup V_c, E')$ such that for vertices $u,v \in V$ and $v_c \in V_c$, where $v_c$ is a copy of vertex $v$, there is an edge $(u,v_c) \in E'$ if and only if $(u,v) \in \vec{E}$.  In the directed graph $G$ the adjacency matrix $A$ is generally asymmetric, however, with our construction we obtain a symmetric adjacency matrix $\mathcal{A}$ for bipartite graph $G'$.
\begin{equation} 
\label{eq:matrixA}
\mathcal{A} =  \left( \begin{array}{cc}
0 & A \\
A^T & 0  \end{array} \right)\,.
\end{equation}

A walk on this undirected bipartite $G'$ starting from a vertex in $V$ will now encounter source nodes in the odd time step and target nodes in the even time step. We call such a walk an \emph{alternating} walk. Formally source and target alternating walks are defined as follows.

\begin{defn} \textbf{The Source Walk.} Given a directed graph, we define \emph{source-walk} of length $k$ as a list of nodes $v_1, v_2, ..., v_{k+1}$ such that there exists edge $(v_i, v_{i+1})$ if $i$ is odd and edge $(v_{i+1}, v_i)$ if $i$ is even:
$v_1 \rightarrow v_2 \leftarrow v_3 \rightarrow \cdots $
\end{defn}

\begin{defn}\textbf{The Target Walk.} A {\em target walk} of length $k$, starting with an in-edge, from node $v_1$ to node $v_{k+1}$ in a directed network is a list of nodes $v_1, v_2, ..., v_{k+1}$ such that there exists edge $(v_{i+1}, v_i)$ if $i$ is odd and edge $(v_i, v_{i+1})$ if $i$ is even:
$v_1 \leftarrow v_2 \rightarrow v_3 \leftarrow \cdots $
\end{defn}
 We now define the alternating random walk which we use to sample the respective neighborhoods of vertices in 3 types of \nerd's computational graphs.
 
 \mpara {Alternating Random Walks.} To generate an alternating random walk we first sample the input node for the source/target walks from the indegree/outdegree distribution of $G$. We then simulate source/target random walks of length $\ell$. Let $c_i$ denote the $i$-th node in the alternating random walk starting with node $u$. 
Then

\[\Pr(c_i=v' | c_{i-1}=v) = \begin{cases}
		{1 \over d^{out}(v)}\cdot w(v,v') , & \text{if } (v,v') \in \vec{E} \\
        {1 \over d^{in}(v)}\cdot w(v',v) , & \text{if } (v',v) \in \vec{E} \\
		0, & \text{otherwise}
	\end{cases}. \]

All nodes in a source/target walk in their respective roles constitute a neighborhood set for the input (the walk starts at the input node) node. 

\begin{figure*}[t]
\centering
    \subfloat[\nerd Alternating Random walks]{\includegraphics[width=0.45\textwidth]{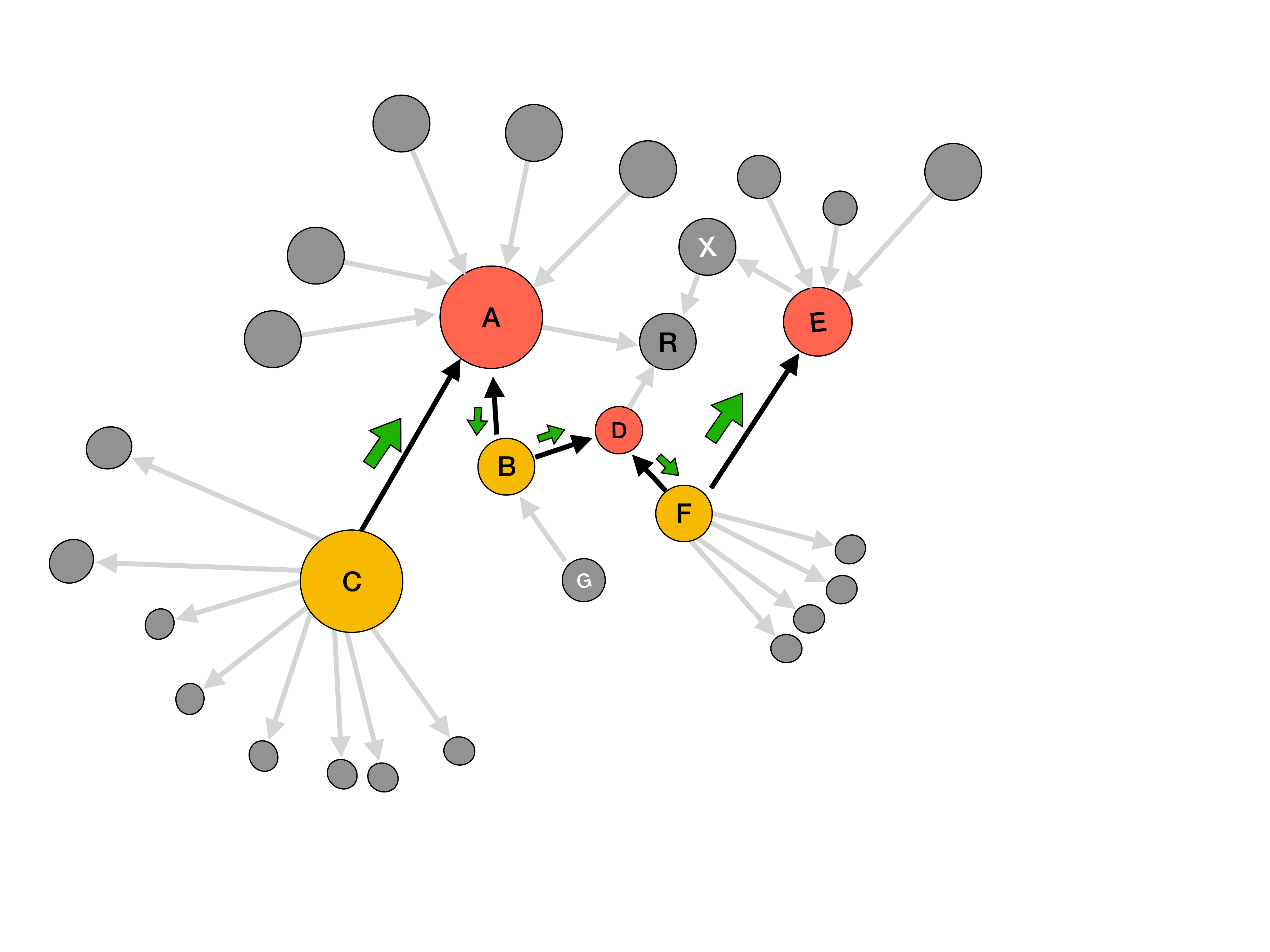}}
    \hfill
    \subfloat[Source and Target  Embeddings in \nerd]{\includegraphics[width=0.45\textwidth]{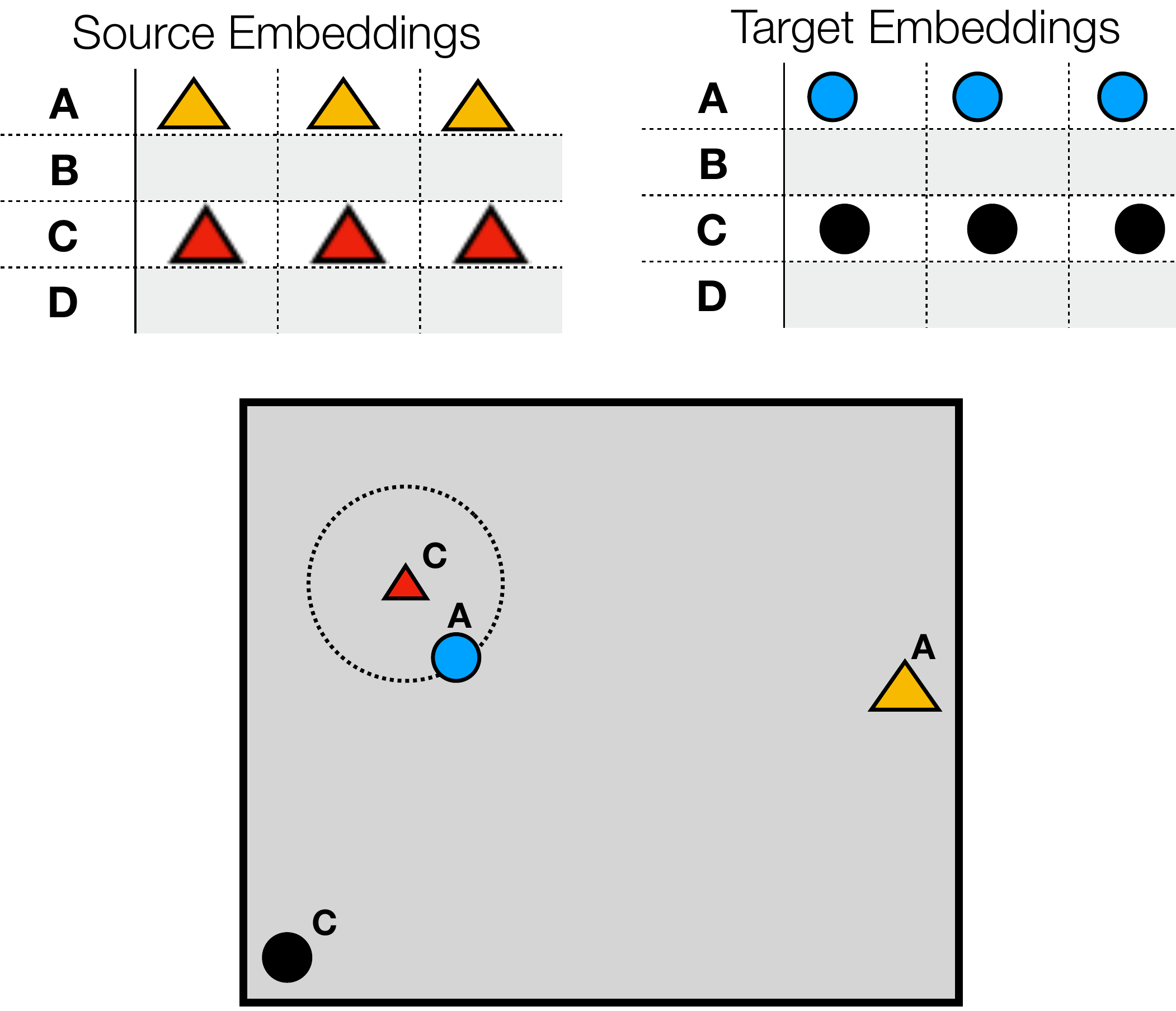}}
\caption{\nerd performing a source walk with input vertex C on an example graph. Node C is the input node. Nodes C, B and F are in source roles and nodes A, D and E are in in target roles. Nodes A, D, E and B,F constitute the target and source neighborhood of source node C. In Figure (b), we show two embedding representations for nodes in their source and target roles. Nodes C and A will be embedded closer in their source-target roles whereas will be far away in source-source or target-target roles. }
\label{fig:nerd-main-fig}
\end{figure*}

\subsection{Learning Framework using Computation Graphs }
\label{sec:learning}
As already mentioned \nerd can be intuitively understood as optimizing first order proximity (embedding vertices sharing an edge closer) in 3 types of computation graphs that it extracts from $G$ via alternating walks. In particular \nerd operates on (i) directed source-target graphs ($G_{st}$) in which a directed edge exists between nodes of opposite roles (ii) source-source  graphs ($G_{s}$) in which an undirected edge between nodes represents the similarity between two nodes in their source roles (iii) target-target graphs ($G_{t}$) in which an undirected edge between nodes represents the similarity between two nodes in their target roles. For example corresponding to a source walk say $v_1 \rightarrow v_2 \leftarrow v_3 \rightarrow \cdots $, $(v_1,v_2)$ form an edge in $G_{st}$, $(v_1,v_3)$ form an edge in $G_{s}$.
For a node/neighbor pair $u,v$ in roles $r_1$ and $r_2$ respectively in any of the computation graphs, we are interested in finding representations $\Phi_{r_1}(u)$ and $\Phi_{r_2}(v)$ in their respective roles such that the following objective is maximized.
 
\begin{align}
\label{eq:obj}
\mathcal{O}(u,v)= & \log \sigma(\Phi_{r_1}(u)\cdot \Phi_{r_2}(v)) + \kappa\mathbb{E}_{v' \sim P^n_{r_2}(v')}(\log\sigma(- \Phi_{r_1}(u)\cdot \Phi_{r_2}(v')),
\end{align}
where $\sigma(x)={1\over 1+ \exp(-x)}$ and $P^n_{r_2}(v')$ is the indegree or outdegree noise distribution and $\kappa$ is the number of negative examples . We set $P^n_{r_2}(v')= {d^{3/4}(v)\over \sum_{v\in V}d^{3/4}(v)}$, where $d$ is the indegree (if $r_2$ is the target role) or outdegree (if $r_2$ is the source role) of vertex $v$. We optimize Equation~\eqref{eq:obj} using Asynchronous Stochastic Gradient Descent \cite{Hogwild}. 

Figure \ref{fig:nerd-main-fig} shows a toy example depicting the working of \nerd. The pseudo-code for \nerd  is stated in Algorithm~\ref{alg:1}. \nerd performs a total of $\gamma$ walks each walk being source walk or target walk with probability $0.5$. The procedure for training a source or target walk is stated in Algorithm \ref{alg:trainWalk}. The first vertex of the walk is the input vertex whose proximity is optimized (using negative samples) with respect to its neighbors in the opposite role (in line 14) and in the same role (in line 16). The joint training with respect to neighbors of same role can be controlled by a binary parameter \textsc{Joint}.

\begin{algorithm}[ht!]
\begin{algorithmic}[1]
 \Require{graph $G(V,\vec{E}_w)$, number of nodes to be sampled of each type $n$, embedding size $d$, number of walks $\gamma$, number of negative samples $\kappa$}, $\textsc{joint} \in \{0,1\} $
\Ensure{matrix of source representations $\Phi_s \in \mathbb{R}^{|V|\times d}$ and target representations $\Phi_t \in \mathbb{R}^{|V|\times d}$}
\Function{\nerd}{$G, n, d,\gamma, \kappa$}

	\State{Initialize $\Phi_s$ and $\Phi_t$}
	\For{$i=0$ \ldots $\gamma$}
          \If{(rand()$~>0.5$)}  
          	\State{ $s_1 \sim P^{out}$ } 
          	\State $W_s = \textsc{SourceWalk}(s_1)$
          	\State\Call{Train}{$W_s,s,\kappa$,\textsc{joint}}~~~~~~~~~~ \Comment{source role s}
          \Else
          \State{ $t_1  \sim P^{in}$ } 
          \State $W_t = \textsc{TargetWalk}(t_1)$
          \State\Call{Train}{$W_t,t,\kappa,\textsc{joint}$} ~~~~~~~\Comment{target role t}
      \EndIf
    \EndFor
\EndFunction

\end{algorithmic}
\caption{NERD}
\label{alg:1}
\end{algorithm}
\begin{algorithm} [h!]
\caption{Train a source or target walk}
\label{alg:trainWalk}
\begin{algorithmic}[1]
    \Function{Train}{$W,r,\kappa,\textsc{joint}$}   
    \State $u\leftarrow W[0]$
    \State $error=0$ 
    \For {$i=1,3,\ldots 2n-1$}
         \For{$j =0\ldots \kappa$}
         \If{$(j=0)$}
           \State $v1= W[i]$      ~~~\Comment{{ neigbor  in opposite role $r'$}}
           \State $v2= W[i+1] $ ~~~\Comment{{ neigbor  of same role $r$}}
           \State $label =1$
           \Else~~~~~~~~~~~~~~~~~~~~~~~~~~~\Comment{ negative samples}
           \State $label=1$
           \State $v1 \sim P_{r'}^n$
           \State $ v2 \sim P_r^n$
		    \EndIf
			\State{$error+=$\Call{Update}{$\Phi_r(u),\Phi_{r'}(v1),label$} }
			\If{(\textsc{joint})}
			\State{$error+=$\Call{Update}{$\Phi_r(u),\Phi_r(v2),label$} }
			\EndIf
		  \EndFor
		\EndFor
		\State{$\Phi_r(u) \,\,+=\,\, error$}
		 \EndFunction
\\
\Function{Update}{$\Phi(u),\Phi(v),label$} ~\Comment{//gradient update}
\State $g = (label - \sigma(\Phi(u)\cdot \Phi(v))\cdot \lambda $ 
\State {$\Phi(v) += g* \Phi(u) $}
\State \Return {$g* \Phi(v)$}
\EndFunction
\end{algorithmic}
\end{algorithm}

We further derive closed form expression for \nerd's optimization in the matrix framework as stated in the following theorem (for proof see Supplementary Material). Note that \nerd (non-joint) refers to when the optimization is done only on the source/target graph.
\begin{theorem} \label{thm:main}
   For any two vertices $i$ and $j$, \nerd (non-joint) finds source ($\Phi_s(i)$) and target ($\Phi_t(j)$) embedding vectors such that $\Phi_s(i)\cdot \Phi_t(j)$ is the $(i,j)th$ entry of the following matrix
  $$\log({vol(G)} \sum_{r\in \{1,3,...,2n-1\}} (D^{-1}\mathcal{A})^r D^{-1})-\log \kappa, $$
   where $n$ and $\kappa$ are as in Algorithm~\ref{alg:1} and $\mathcal{A}$ is the adjacency matrix of the bipartite network $G'$ obtained by mapping the given directed network $G$ to $G'$ as defined in \eqref{eq:matrixA}.
\end{theorem}

\mpara{Complexity Analysis.} Sampling a vertex based on indegree or outdegree distribution requires constant amortized time by building an alias sampling table upfront. At any time only $2n$ neighbors are stored which is typically a small number as we observed in our experiments. In our experiments we set the total number of walks  equal to $800$ times the number of vertices. In each optimization step we use $\kappa=\{3,5\}$ negative samples, therefore, complexity of each optimization step is $O(d\kappa)$. As it requires $O(|E|)$ time to read the graph initially, the run time complexity for \nerd can be given as $O(n d \kappa N+ |E|)$. The space complexity of \nerd is $O(|E|)$. As our method is linear (with respect to space and time complexity) in the input size, it is scalable for large graphs.
\section{Experiments}
 In this section we present how we evaluate \nerd\footnote{We make available our implementation with corresponding data at \url{https://git.l3s.uni-hannover.de/khosla/nerd}} against several state-of-the-art node embedding algorithms. We use the original implementations of the authors for all the baselines (if available). We also employ parameter tuning whenever possible (parameter settings are detailed in Supplementary Material), otherwise we use the best parameters reported by the authors in their respective papers. We perform comparisons corresponding to three tasks -- Link Prediction, Graph Reconstruction and Node classification. In the next section we explain the datasets used in our evaluations.

\begin{table}
\setlength{\tabcolsep}{3.5pt}
\centering

\small
\begin{tabular}{lcrccrccccr}
\multicolumn{1}{c}{} & \multicolumn{3}{c}{\textbf{Size}} & \multicolumn{2}{c}{\textbf{Statistics}}
& \multicolumn{3}{c}{\textbf{Details}}\\
\cmidrule(lr){2-4}\cmidrule(lr){5-6}\cmidrule(lr){7-9}
\emph{dataset} & \(|V|\) & \multicolumn{1}{c}{\(|E|\)} & \(|\mathcal{L}|\) & \mbox{Diameter} & \mbox{Reciprocity} & \mbox{Labels} & \mbox{vertex} & \mbox{edges} \\
    \midrule

\cora & 23,166 & 91,500 & 79 & 20 & 0.051 & \checkmark & papers & citation \\
\twitter & 465,017& 834,797 & - & 8 & 0.003 & - & people & follower \\ 
\epi & 75,879 & 508,837 & - & 15 & 0.405 & - & people & trust\\ 
\pubmed& 19,718 & 44,327 & 3 & 18 & 0.0007 & \checkmark & papers & citation \\ 
\cocit&44,034 & 195,361 & 15 & 25 & 0 & \checkmark & papers & citation\\ 

\bottomrule
\end{tabular}
\caption{Dataset characteristics: number of nodes \(|V|\), number of edges \(|E|\); number of node labels \(|\mathcal{L}|\).}
\label{tbl:datasets}
\vspace{-5mm}
\end{table}

A brief summary of the characteristics of the datasets (details are links provided in the implementation page) is presented in Table~\ref{tbl:datasets}. We recall that \emph{reciprocity} in a directed graph equals the proportion of edges for which an edge in the opposite direction exists, i.e., that are reciprocated. All the above datasets except PubMed and Cocitation datasets have been collected from \cite{kunegis2015konect}.

\mpara{Baselines.}
We compare the \nerd model with several existing node embedding models for link prediction, graph reconstruction and node classification tasks.
As baselines we consider methods like \app \cite{AAAI1714696} and \hope~\cite{ou2016asymmetric} which uses two embedding spaces to embed vertices in their two roles. We also compare against other popular single embedding based methods: \deepwalk~\cite{Perozzi:2014}, \lineemb~\cite{tang2015line}, \nodetovec~\cite{Grover:2016} and \verseemb~\cite{tsitsulin2018verse}. All these methods are detailed in related work. As \nerd is an unsupervised shallow neural method, for fair comparisons, semi-supervised or unsupervised deep models are not considered as baselines.

\subsection{Link Prediction}

The aim of the link prediction task is to predict missing edges given a network with a fraction of removed edges. 
A fraction of edges is removed randomly to serve as the \emph{test split} while the residual network can be utilized for training. 
The test split is balanced with negative edges sampled from random vertex pairs that have no edges between them. We refer to this setting as the \emph{undirected} setting. 
While removing edges randomly, we make sure that no node is isolated, otherwise the representations corresponding to these nodes can not be learned.


\mpara{Directed link prediction.} 
Since we are interested in not only the existence of the edges between nodes but also the directions of these edges, we consider a slight modification in the test split setting. 
Note that this is a slight departure from the experimental settings used in previous works where only the presence of an edge was evaluated. 
We posit that in a directed network the algorithm should also be able to decide the direction of the predicted edge. 
To achieve this, we allow for negative edges that are complements of the true (positive) edges which exist already in the test split.

We experiment by varying the number of such complement edges created by inverting a fraction of the true edges in the test split. 
A value of $0$ corresponds to the classical undirected graph setting while a value in $(0, 1]$ determines what fraction of positive edges from the test split are inverted at most to create negative examples. It can also happen that an inverted edge is actually an edge in the network, in which case we discard it and pick up some random pair which corresponds to a negative edge. Such a construction of test data is essential to check if the algorithm is also predicting the correct direction of the edge along with the existence of the edge. 
Please note that we always make sure that in the test set the number of negative examples is equal to the number of positive examples. Embedding dimensions are set to $128$ for all models for both settings.

Table \ref{tab:directed-test} presents the ROC-AUC (Area Under the Receiver Operating Characteristic Curve) scores for link prediction for three datasets (missing datasets show similar trends, results not presented because of space constraints). More specifically, given an embedding, the inner product of two node representations normalized by the sigmoid function is employed as the similarity/link-probability measurement for all the algorithms. Fraction $0\%$ correspond to the undirected setting in which the negative edges in the test set are randomly picked. The $50\%$ and $100\%$ corresponds to directed setting in which at most $50\%$ and $100\%$ positive edges of test set are inverted to form negative edges. Please note that if an inverted edge is actually an edge in the network, we discard it and pick up some random pair.

\mpara{Performance on Cora.}
\verseemb outperforms others for the undirected setting in the Cora dataset. But its performance decreases rapidly in the directed setting where the algorithm is forced to assign a direction to the edge.
The performance of the three directed methods (\app, \hoppe and \nerd) is stable supporting the fact that these methods can correctly predict the edge direction in addition to predicting a link.
\nerd is the next best (AUC of $0.788$) and outperforms \hoppe for directed setting with $50\%$ and $100\%$ (AUC of $0.813$) test set edge reversal. 
This means that that whenever \nerd predicts the presence of an edge it in fact also predicts the edge directionality accurately.

\begin{figure}[h!]

\centering
    \includegraphics[width=0.6\linewidth]{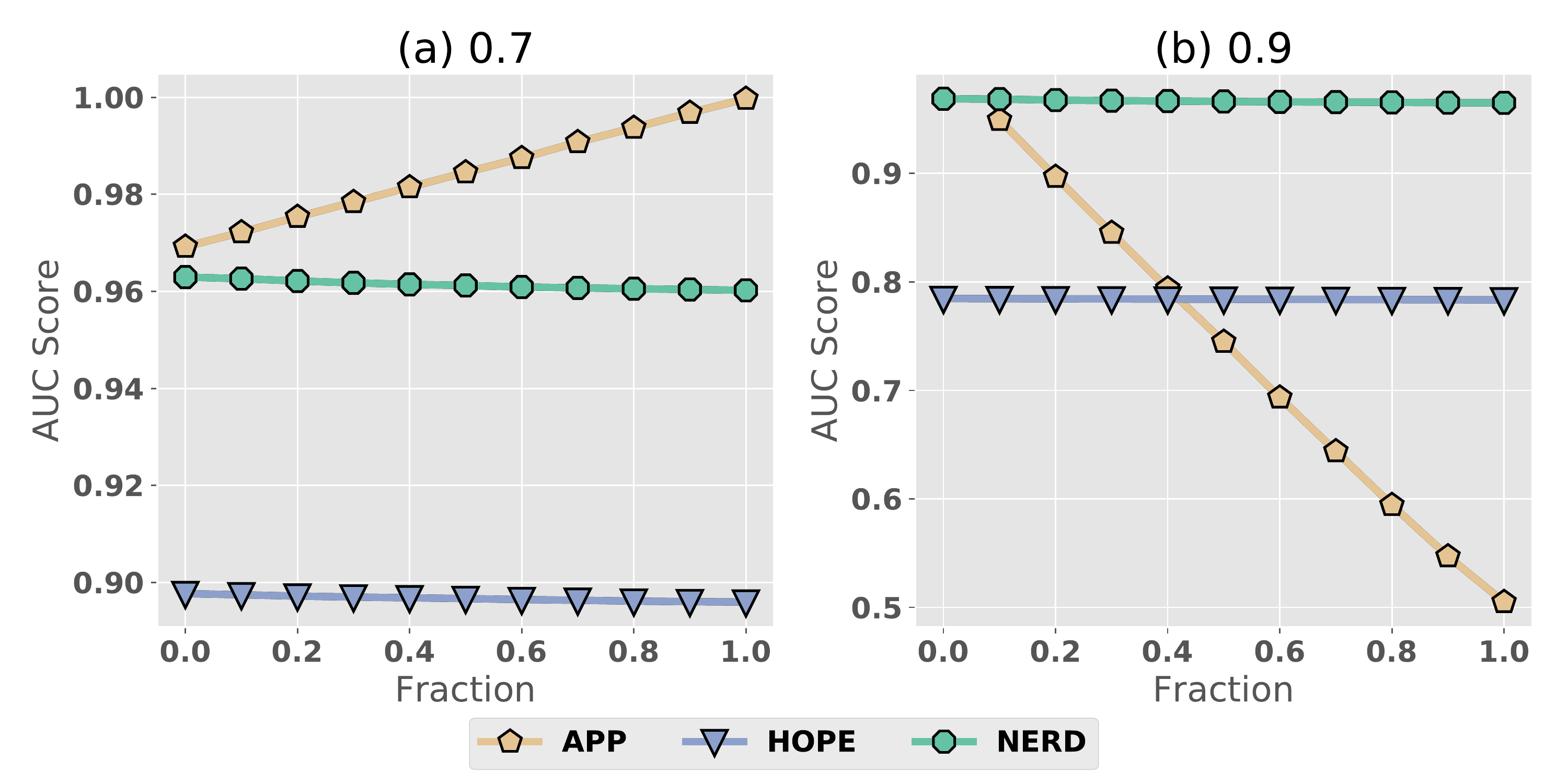}
\caption{Link prediction in Twitter. The y-axis shows the AUC scores and the x-axis is the maximum fraction of edges that are inverted in the test split. The models are trained on $70\%$ and $90\%$ of the Twitter edges respectively. The fraction on the x-axis indicates the maximum fraction of inverted positive test edges to create negative test edges. Note that the train-test split is the same over all fractions.
}
\label{fig:lp-cora-twitter}

\end{figure}

 \mpara{Performance on Twitter.} For the Twitter dataset, \hoppe outperforms all other methods and is closely followed by \nerd for 60-40 split of training-test data. 
 Figure \ref{fig:lp-cora-twitter} shows the performance of three directed graph methods: \app, \hoppe and \nerd on 70-30 and 90-10 training-test splits for Twitter respectively. 
 Here we plot the AUC scores by varying the fraction of inverted edges in the test split to construct negative test edges. 
We omit other methods as all of them have a very low performance.
 We make several interesting observations here. First, \hoppe which performs best for 60-40 split shows a decrease in performance with the increase in training data. 
 We believe that the parameters for \hoppe namely the attenuation factor which was tuned for best performance on a smaller amount of training data no longer might not be applicable for larger training data. This renders such a method to be very sensitive to structural changes in the graph. 
 Second, \app's performance improves with increasing training data but is not as stable as \nerd and \hoppe in the directed setting when the fraction of inverted edges is increased, i.e., it does not always correctly predict the direction of an edge.
 Third, \nerd's performance stays stable and improves on increasing the training data, which confirms our justification that it is more robust to structural changes caused by random addition/removal of edges. 
 Moreover, at $90\%$ training data it is the best performing method and second best but consistent (in predicting edge direction) for other splits.
 Finally we observe that Twitter has a prominent hub-authority structure with more than than $99\%$ vertices with zero out-degree. Using non-alternating directed walks on Twitter hinders \app and other similar random walk methods to fully explore the network structure as much as they could do for Cora. 

\mpara{Performance on Epinions.} \verseemb shows a high performance on Epinions in undirected setting which is not surprising as Epinions has a high reciprocity with more than $40\%$ of the edges existing in both directions. 
\nerd on the other hand beats the two other directed methods \app and \hoppe for both the settings.  As the fraction of edge reversals increases, \nerd also starts performing better than \verseemb. 
We note that even though \nerd does not outperforms all methods on link prediction, it shows more robustness across datasets being the second best performing (when not the best) and is consistent in predicting the right edge direction i.e., its performance does not vary a lot (except in Epinions with high reciprocity) with increasing fraction of positive test edge inversions in the directed setting. 

\begin{table*}[ht!!]
\begin{center}
\setlength{\tabcolsep}{1pt}
\renewcommand{\aboverulesep}{0pt}
\renewcommand{\belowrulesep}{0pt}
\renewcommand\arraystretch{1.35}
\newcolumntype{C}{>{\centering\arraybackslash}X}

\begin{tabularx}{\textwidth}{p{2.6cm}|CCC|CCCHHH|CCC}
\multicolumn{1}{c}{} & \multicolumn{3}{c}{\cora}& \multicolumn{3}{c}{\twitter}&\multicolumn{3}{H}{}&\multicolumn{3}{c}{\epi}\\
\cmidrule{2-4}  \cmidrule{5-7}  \cmidrule{8-10}\cmidrule{11-13}
\multicolumn{1}{l}{\emph{method}} & 0\% & 50\% & 100\% & 0\% & 50\% & 100\% & 0\% & 50\% & 100\% & 0\% & 50\% & 100\% \\
\hline
\rowcolor{lightgray}
\small\deepwalk & 0.836 & 0.669 & 0.532 & 0.536 & 0.522 & 0.501 & 0.868 & 0.680& 0.503 & 0.538 & 0.560 & 0.563 \\

\small\nodetovec  &0.840 & 0.649 & 0.526 &  0.500&  0.500 & 0.500 & 0.889 & 0.697&0.503 & \underline{0.930} & 0.750 & 0.726 \\

\rowcolor{lightgray}
\small\verseemb  & \textbf{0.875} & 0.688 & 0.500 &0.52& 0.510 & 0.501 & 0.809 & 0.654 & 0.503 & \textbf{0.955} & \underline{0.753} & \underline{0.739} \\

\small\app  & \underline{0.865} & \bf{0.841}& \bf{0.833} & 0.723& 0.638 & 0.555 & \textbf{0.957} & \textbf{0.838}& 0.722 & 0.639 & 0.477 & 0.455 \\

\rowcolor{lightgray}
\small\hoppe  & 0.784 & 0.734 & 0.718 & \textbf{0.981}& \textbf{0.980} & \textbf{0.979} & 0.756 & 0.737& \textbf{0.732} & 0.807 & 0.718 & 0.716\\

\small\lineemb-{1+2}  & 0.735 & 0.619 & 0.518 & 0.009 & 0.255 & 0.500 & 0.319 & 0.404 & 0.501 & 0.658 & 0.622 & 0.617 \\

\rowcolor{lightgray}
\small\lineemb-{1} & 0.781 & 0.644 & 0.526  & 0.007 &  0.007 &  0.254 & 0.312 & 0.405 & 0.501 & 0.744 & 0.677 & 0.668 \\

\small\lineemb-{2}  & 0.693 & 0.598 & 0.514 & 0.511 & 0.507 & 0.503 & 0.642 & 0.572 & 0.503 & 0.555 & 0.544 & 0.543 \\
\hline
\small\nerd  &0.795  &\underline{0.788}& \underline{0.813} & \underline{0.969} & \underline{0.968} & \underline{0.967} & && &0.906& \bf{0.774} & \bf{0.771}  \\
\bottomrule
\end{tabularx}
\end{center}

\caption{Link Prediction Results for directed graphs with (1) random negative edges in test set (2) 50\% of the test negative edges created by reversing positive edges (3) when all positive edges are reversed to create negative edges in the test set. The top scores are shown in bold whereas the second best scores are underlined. }
\label{tab:directed-test}
\vspace{-0.8cm}
\end{table*}

\subsection{Graph Reconstruction}
\label{sec:gr}
In the graph reconstruction task we evaluate how well the embeddings preserve neighborhood information. There are two separate evaluation regimes for graph reconstruction in previous works. One line of work~\cite{ou2016asymmetric}, that we refer to as \emph{edge-centric} evaluation, relies on sampling random pairs of nodes from the original graphs into their test set. These candidate edges are then ordered according to their similarity in the embedding space. Precision is then computed at different rank depths where the relevant edges are the ones present in the original graph. On the other hand,~\cite{tsitsulin2018verse} perform a \emph{node-centric} evaluation where precision is computed on a per-node basis. For a given node $v$ with an outdegree $k$, embeddings are used to perform a $k$-nearest neighbor search for $v$ and precision is computed based on how many actual neighbors the $k$-NN procedure is able to extract.

\mpara{Directed Graph Reconstruction.} We believe that the edge-centric evaluation suffers from sparsity issues typical in real-world networks and even if a large number of node pairs are sampled, the fraction of relevant edges retrieved tends to remain low. More acutely, such an approach does not model the neighborhood reconstruction aspect of graph construction and is rather close to predicting links. We adopt the node-centric evaluation approach where we intend to also compute precision on directed networks with a slight modification. In particular, we propose to compute precision for both outgoing and incoming edges for a given node. This is different from previous evaluation approaches which only considers the reconstruction of adjacency list of a node, i.e., only its outgoing neighbors. Moreover in our proposed evaluation strategy we do not assume the prior knowledge of the indegree or outdegree.

As in Link Prediction, the similarity or the probability of an edge $(i,j)$ is computed as the sigmoid over the dot product of their respective embedding vectors. For \hoppe, \nerd and \app we use the corresponding source and target vectors respectively.
We do not assume the prior knowledge of the indegree or outdegree, rather we compute the precision for $k\in\{1,2,5,10,100,200\}$. For a given $k$ we obtain the $k$-nearest neighbors ranked by sigmoid similarity for each embedding approach. If a node has an outdegree or indegree of zero, we set the precision to be $1$ if the sigmoid corresponding to the nearest neighbor is less than $0.51$ (recall that $\sigma(\vec{x} \cdot \vec{y}) = 0.5$ for $\vec{x} \cdot \vec{y} = 0$), otherwise we set it to $0$. In other cases, for a given node $v$ and a specific $k$ we compute ${P^k_{out}(v)}$ and ${P^k_{in}(v)}$ corresponding to the outgoing and incoming edges as
$$ P^k_{out}(v) = { \mathcal{N}^k_{out} \cap N^{out}(v) \over k},~~~ P^k_{in}(v) = { \mathcal{N}^k_{in} \cap N^{in}(v) \over k},$$
where $\mathcal{N}^k_{out}(v)$ and $\mathcal{N}^k_{in}(v)$ are the $k$ nearest outgoing (to whom $v$ has outgoing edges) and incoming (neighbors point to $v$) neighbors retrieved from the embeddings and $N^{out}(v)$ and $N^{in}(v)$ are the actual outgoing and incoming neighbors of $v$.
We then compute the Micro-F1 score as the harmonic mean of $P^k_{in}(v)$ and $P^k_{out}(v)$.
To avoid any zeros in the denominator, we add a very small $\varepsilon=10^{-5}$ to each precision value before computing the harmonic mean.
We finally report the final precision as the average of these harmonic means over the nodes in the test set. 

\begin{figure*}[h!]
\centering
    \includegraphics[width=0.8\linewidth]{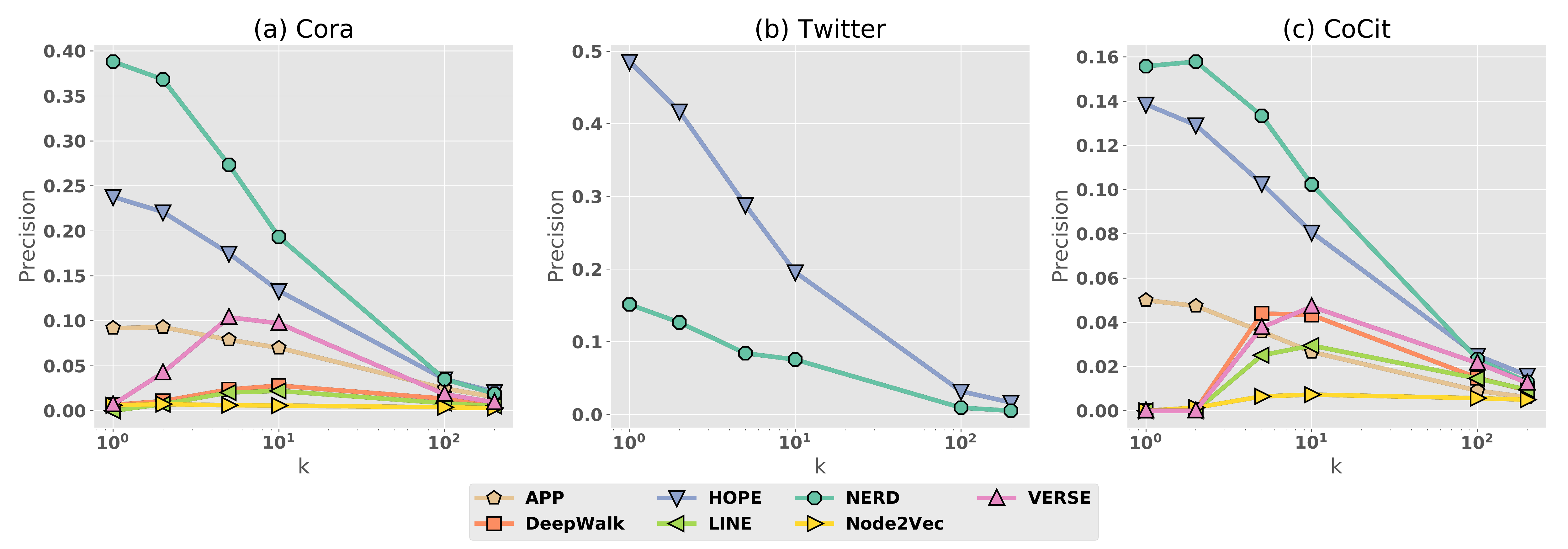}

\caption{Graph reconstruction for Cora, Twitter and CoCitation Networks with precision computed on both outgoing and incoming edges. \nerd shows an improvement of $63.1\%$ (for $k=1$) as compared to \hoppe in the Cora dataset. The trend is reversed in the Twitter dataset because of the exact nature of \hoppe on low-rank Twitter data. For Twitter, all methods except \nerd and \hoppe have precision close to $10^{-5}$, therefore we do not show them in the plots. \nerd shows an improvement of $12.5\%$ (for $k=1$) as compared to \hoppe in the CoCit dataset.}
\vspace{-1mm}
\label{fig:gr-plots}
\end{figure*}

\mpara{Results.} We perform the graph reconstruction task on the Cora, Cocitation and Twitter datasets. In order to create the test set we randomly sample $10\%$ of the nodes for Cora and Cocitation datasets and and $1\%$ of Twitter. We plot the final averaged precision corresponding to different values of $k$ in Figure~\ref{fig:gr-plots}. 

For the Cora dataset, \nerd clearly outperforms all the other models including \hoppe. In particular for $k=1$, \nerd shows an improvement of $63\%$ over \hoppe which in some sense is fine tuned for this task. 

 The trend between \nerd and \hoppe is reversed for Twitter dataset, where \hoppe behaves like an almost exact algorithm. This can be attributed to the low rank of the associated Katz similarity matrix. Note that only $2502$ out of more than $400K$ nodes have non-zero outdegree which causes a tremendous drop in the rank of the associated Katz matrix. We recall that \hoppe's approximation guarantee relies on the low rank assumption of the associated similarity matrix which seems to be fulfilled quite well in this dataset. The performance of other models in our directed setting clearly shows their inadequacy to reconstruct neighborhoods in directed graphs. For Twitter, we only show plots corresponding to \hoppe and \nerd as precision corresponding to other methods is close to $10^{-5}$. 

Again for Cocitation \nerd performs the best with an improvement of around $12.5\%$ for $k=1$ over the second best performing method, \hoppe.
Once again, \nerd exhibited robustness in this task as for Twitter, it is closest to the best performing method. 
 Note that some of the methods like \verseemb and \app which were sometimes better performing than \nerd in link prediction show a poor performance across all datasets in graph reconstruction task. 
 Note that this task is harder than link prediction as the model not only needs to predict the incoming and outgoing neighbors but also has no prior knowledge of the number of neighbors. 
 Moreover the test set for this task is not balanced in the sense that for each test node the model needs to distinguish between small number of positive edges with a huge number of negative edges, for example for small $k$.

\begin{table*}[h!]
\begin{center}
\setlength{\tabcolsep}{1pt}
\renewcommand{\aboverulesep}{0pt}
\renewcommand{\belowrulesep}{0pt}
\renewcommand\arraystretch{1.35}

\newcolumntype{C}{>{\centering\arraybackslash}X}

\begin{tabularx}{\textwidth}{p{2.8cm}|HHCC|CCHHHHHH|CC}

\multicolumn{1}{p{2.8cm}}{} & \multicolumn{2}{H}{}& \multicolumn{2}{c}{\pubmed} & \multicolumn{2}{c}{\cora} & \multicolumn{2}{H}{} & 
\multicolumn{2}{H}{} & 
\multicolumn{2}{H}{}& \multicolumn{2}{c}{\cocit}\\
\cmidrule{2-15}

\multicolumn{1}{l}{\emph{method}} & mic. & mac. & mic. & mac. & mic. & mac. & mic. & mac. & mic. & mac. & mic. & mac. & mic. & mac.\\
\midrule

\deepwalk & & & 73.96 &71.34 & 64.98&\bf{51.53} &94.40&92.01 & & \textbf{31.00}& &39.89& 41.92&30.07 \\

\rowcolor{lightgray}
\small
\nodetovec{}  &42.46&29.16& 72.36&68.54& 65.74&49.12 & 94.11&91.73& 	42.11&30.57&48.41&42.04 & 41.64&28.18 \\

\small
\verseemb{} &  35.51&21.77 & 71.24&68.68  & 60.87&45.52& 92.87&89.69&  35.70 &	23.00 &45.12&37.28 & 40.17 & 27.56 \\

\rowcolor{lightgray}
\small
\app{} & 20.60&5.39& 69.00&65.20 & 64.58&47.03&  77.11&56.28 & 24.26&4.21 & 45.04&36.61 & 40.34&28.06 \\

\small 
\hoppe  & n.a & n.a & 63.00 & 54.6 & 26.23 & 1.22  & n.a & n.a & n.a & n.a & n.a & n.a & 16.66 & 1.91 \\

\rowcolor{lightgray}
\small
\lineemb-{1+2}  & 41.01&25.02 &  62.29&59.79& 54.04&41.83 &\textbf{94.50}&92.08 & 41.46&27.65 &48.22&41.51&37.71 & 26.75  \\

\small
\lineemb-{1}  & 41.54&24.28 &  55.65&53.83& 62.36&47.19 &94.31&91.96 & 40.92&26.19 & 47.49&41.17&36.10&25.70  \\

\rowcolor{lightgray}
\small
\lineemb-{2}  & 36.70&18.80 &  56.81&51.71& 51.05&35.37&94.30&91.81 & 40.49&24.24 & 47.46&39.97&31.4&20.59  \\

\hline
\small
\nerd  & &&\bf{76.70} & \bf{74.67}  & \bf{67.75} & \underline{51.30}&&& && &&\bf{44.84} & \bf{33.49}  \\

\bottomrule
\end{tabularx}
\end{center}
\caption{Node Classification results in terms of Micro-F1 and Macro-F1. All results are mean of 5-fold cross validations. Each node in Cora has multiple labels, for PubMed and CoCit, there is a single label per node. }\label{tab:node-classification}
\vspace{-5mm}
\end{table*}

\subsection{Node Classification}
We run experiments for predicting labels in the Cora, CoCitation and PubMed datasets (labels were not available for other two datasets). We report the Micro-F1 and Macro-F1 scores after a $5$-fold multi-label classification using one-vs-rest logistic regression. The main aim of this experiment is to show that \nerd is generalizable across tasks and also performs well for a task like node classification which is not fine tuned for directed graphs. Unlike \app and \hoppe, \nerd also performs the best in this task over all the 3 datasets. Other single embedding based methods like \deepwalk and \nodetovec also exhibit a good performance for node classification indicating that edge directionality might not be a very important factor for node labels at least for the studied datasets. \hoppe which performs very well for link prediction and graph reconstruction tasks shows a poorer performance. 

As we already pointed out that \hoppe is tied to  particular proximity matrix and adjusting it for a task becomes much harder and non-intuitive than random walk based methods where hyper parameters can be easily fine-tuned. 
We also note that for \hoppe the similarity between nodes $i$ and $j$ is determined by the effective distance between them which is computed using the Katz measure, penalizing longer distances by an attenuation factor $\beta$. The advantage of such a degrading distance measure is that it conserves the adjacency similarity of the graph, which reflects in our experiments on Graph Reconstruction. \nerd on the other hand also takes into account how likely $i$ can influence $j$ by taking into account the likelihood of the traversal of various alternating paths between $i$ and $j$. In other words, \nerd constructs the local neighborhood based on how likely this neighborhood can influence the node, which helps the classifier to learn better labels on \nerd trained embeddings. 

\section{Conclusion}

We presented a novel approach, \nerd, for embedding directed graphs while preserving the role semantics of the nodes. We propose an alternating random walk strategy to sample node neighborhoods from a directed graph. The runtime and space complexities of \nerd are both linear in the input size, which makes it suitable for large scale directed graphs.
In addition to providing advantages of using two embedding representations of nodes in a directed graph, we revisit the previously used evaluation strategies for directed graphs. We chart out a clear evaluation strategy for link prediction and graph reconstruction tasks. 
We observe in our experiments where we find that a method performing best in one of the tasks might perform the worst in the other task. 
This beats the whole idea of unsupervised learning which is supposed not to be fine tuned towards a particular task but should be performing well across different tasks. We show that the embeddings from \nerd are indeed robust, generalizable and well performing across multiple types of tasks and networks. We have showcased the effectiveness of \nerd by employing a shallow neural model to optimize the topology of the extracted computational graphs. In future we will employ deeper models to capture non-linearities while preserving the respective topologies.
\vspace{-5mm}
\subsubsection*{Acknowledgements}
\small{This work is partially funded by SoBigData (EU's Horizon 2020  grant agreement No. 654024).} 
\vspace{-5mm}

\bibliographystyle{splncs04}

\begin{thebibliography}{10}

\bibitem{learnEdge2017}
Abu-El-Haija, S., Perozzi, B., Al-Rfou, R.: Learning edge representations via
  low-rank asymmetric projections. pp. 1787--1796. CIKM '17 (2017)

\bibitem{ammann1993robust}
Ammann, L.P.: Robust singular value decompositions: A new approach to
  projection pursuit. Journal of the American Statistical Association
  \textbf{88}(422) (1993)

\bibitem{NIPS2001_1961}
Belkin, M., Niyogi, P.: Laplacian eigenmaps and spectral techniques for
  embedding and clustering. In: Advances in Neural Information Processing
  Systems 14, pp. 585--591 (2002)

\bibitem{cao2015grarep}
Cao, S., Lu, W., Xu, Q.: Grarep: Learning graph representations with global
  structural information. In: CIKM '15. pp. 891--900 (2015)

\bibitem{Cao:2016}
Cao, S., Lu, W., Xu, Q.: Deep neural networks for learning graph
  representations. In: Proceedings of AAAI. pp. 1145--1152. AAAI'16 (2016)

\bibitem{chen2007directed}
Chen, M., Yang, Q., Tang, X.: Directed graph embedding. In: IJCAI(2007)

\bibitem{drobyshevskiy2017learning}
Drobyshevskiy, M., Korshunov, A., Turdakov, D.: Learning and scaling directed
  networks via graph embedding. In: ECML-PKDD. pp. 634--650 (2017)

\bibitem{Grover:2016}
Grover, A., Leskovec, J.: Node2vec: Scalable feature learning for networks. In:
  KDD. pp. 855--864. (2016)

\bibitem{katz1953new}
Katz, L.: A new status index derived from sociometric analysis. Psychometrika
  \textbf{18}(1),  39--43 (1953)

\bibitem{kipf2016variational}
Kipf, T.N., Welling, M.: Variational graph auto-encoders. In: NeurIPS Workshop
  on Bayesian Deep Learning (NeurIPS-16 BDL) (2016)

\bibitem{Kleinberg:1999}
Kleinberg, J.M.: Authoritative sources in a hyperlinked environment. J. ACM
  \textbf{46}(5),  604--632 (1999)

\bibitem{kunegis2015konect}
Kunegis, J.: Konect datasets: Koblenz network collection.
  http://konect.uni-koblenz.de  (2015)

\bibitem{lempel2001salsa}
Lempel, R., Moran, S.: Salsa: the stochastic approach for link-structure
  analysis. ACM Transactions on Information Systems (TOIS)  \textbf{19}(2),
  131--160 (2001)

\bibitem{liben2007link}
Liben-Nowell, D., Kleinberg, J.: The link-prediction problem for social
  networks. Journal of the American society for information science and
  technology  \textbf{58}(7),  1019--1031 (2007)

\bibitem{mikolov2013distributed}
Mikolov, T., Sutskever, I., Chen, K., Corrado, G.S., Dean, J.: Distributed
  representations of words and phrases and their compositionality. In:
  Proceedings of the 27th Annual Conference on Neural Information Processing
  Systems 2013. pp. 3111--3119 (2013)

\bibitem{Mousazadeh:2015}
Mousazadeh, S., Cohen, I.: Embedding and function extension on directed graph.
  Signal Process.  \textbf{111}(C),  137--149 (2015)

\bibitem{ou2016asymmetric}
Ou, M., Cui, P., Pei, J., Zhang, Z., Zhu, W.: Asymmetric transitivity
  preserving graph embedding. In: KDD. pp.
  1105--1114. (2016)

\bibitem{arga}
Pan, S., Hu, R., Long, G., Jiang, J., Yao, L., Zhang, C.: Adversarially
  regularized graph autoencoder for graph embedding. In: IJCAI-18. pp.
  2609--2615 (2018)

\bibitem{Perozzi:2014}
Perozzi, B., Al-Rfou, R., Skiena, S.: Deepwalk: Online learning of social
  representations. In: Proceedings of SIGKDD. pp. 701--710. KDD '14 (2014)

\bibitem{perrault2011directed}
Perrault-Joncas, D.C., Meila, M.: Directed graph embedding: an algorithm based
  on continuous limits of laplacian-type operators. In: Advances in Neural
  Information Processing Systems. pp. 990--998 (2011)

\bibitem{Hogwild}
Recht, B., Re, C., Wright, S., N., F.: Hogwild: A lock-free approach to
  parallelizing stochastic gradient descent. In: Advances in Neural Information
  Processing Systems 24, pp. 693--701 (2011)

\bibitem{tang2015line}
Tang, J., Qu, M., Wang, M., Zhang, M., Yan, J., Mei, Q.: Line: Large-scale
  information network embedding. In: Proceedings of the 24th International
  Conference on World Wide Web. pp. 1067--1077 (2015)

\bibitem{tsitsulin2018verse}
Tsitsulin, A., Mottin, D., Karras, P., M{\"u}ller, E.: Verse: Versatile graph
  embeddings from similarity measures. In: {Proceedings of the 2018 World Wide
  Web Conference}. pp. 539--548 (2018)

\bibitem{wang2016structural}
Wang, D., Cui, P., Zhu, W.: Structural deep network embedding. In: Proceedings
  of the 22Nd ACM SIGKDD International Conference on Knowledge Discovery and
  Data Mining. pp. 1225--1234. KDD '16 (2016)

\bibitem{ying2018graph}
Ying, R., He, R., Chen, K., Eksombatchai, P., Hamilton, W.L., Leskovec, J.:
  Graph convolutional neural networks for web-scale recommender systems. In:
  KDD. pp. 974--983 (2018)

\bibitem{AAAI1714696}
Zhou, C., Liu, Y., Liu, X., Liu, Z., Gao, J.: Scalable graph embedding for
  asymmetric proximity. In: AAAI Conference on Artificial Intelligence
  (AAAI'17) (2017)

\end{thebibliography}
 
\begin{appendix}
\section{Missing Proofs}
We further support our approach by deriving a closed form expression for \nerd's optimization in the matrix framework.
We remark that our analysis applies to \nerd's framework when the optimization is only performed over node pairs which have the opposite roles, i.e., operation in line 16 in Algorithm 2 is not performed.

Following the work in \cite{Levy:2014}, for a given training sample of a word $w$ and a context $c$, SGNS learns respective word and context representations $\vec{w}$ and $\vec{c}$ such that
\begin{equation} \label{eq:sgns} \vec{w}\cdot \vec{c}=\log {P(w,c) \over P(w) P(c)} -\log \kappa,\end{equation}
where $P(w,c)$ is the joint probability distribution of $(w,c)$ pairs (occurring in a contextual window) and $P(w)$ and $P(c)$ are the probability distributions of sampled words and contexts respectively and $k$ is the number of negative samples.
As we also employ the SGNS objective for optimization the training pairs which are in opposite roles (here we ignore the training of pairs with same role), we use the main result from \cite{Levy:2014} which implies that for a training pair $(s,t)$ \nerd finds source and target vectors obeying Equation \eqref{eq:sgns} with word and context replaced by the node pairs. To compute the right hand side of \eqref{eq:sgns} we need to compute the  distributions for sampling a training pair and the marginalized node distributions, which we accomplish in the following proof of Theorem 1.
\begin{proof}[Proof of Theorem 1]
We recall that $D=diag(d_1,d_2,...,d_{2N})$ is the degree matrix of $G'$. Set $\mathbf{P}=D^{-1}\mathcal{A}.$ 
First we note that the initial vertex $v$ is chosen with probability $d^{out}(v)\over vol(G)$ for a source walk or $d^{in}(v)\over vol(G)$ for a target walk, starting at $v$. 
The probability that a given source-target pair $(i,j)$ will be sampled in a walk of length $\ell = 2n - 1$, where $n$ is the number of sampled pairs, is given by 
\begin{equation} \label{eq:prob}
P(i,j) =  \sum_{r\in \{1,3,...,2n-1\}} \bigg( {1\over 2} {d(i) \over vol(G')} \cdot (\mathbf{P^{r}})_{i,j} + {1\over 2}{d(j) \over vol(G')} \cdot (\mathbf{P^{r}})_{j,i}\bigg).
\end{equation}
The first term corresponds the sampling of $(i, j)$ in a source walk starting from the source node $i$ and the second term corresponds to the sampling of $(j, i)$ in a target walk starting from the target node $j$. Note that $d(i)$ corresponds to the out-degree of $i$ in the original graph $G$, $d(j)$ corresponds to the in-degree of $j$ in the original graph $G$, and $vol(G) = vol(G')$. Also note that for \nerd the input vertex is always the first vertex in the walk. Further marginalization of Equation \eqref{eq:prob} gives us $P(i) = {d(i) \over vol(G')}$ and $P(j) = {d(j) \over vol(G')}$. From \eqref{eq:sgns} we have $\Phi_s(i)\cdot \Phi_t(j)={P(i,j)\over P(i)P(j)}$ , therefore substitution the above terms we obtain 
\begin{align} \label{eq:intp}
&\Phi_s(i)\cdot \Phi_t(j) =\\
& {\sum_{r\in \{1,3,\cdots,2n-1\}} ( {d(i) \over vol(G')} \cdot (\mathbf{P^{r}})_{i,j} + {d(j) \over vol(G')} \cdot (\mathbf{P^{r}})_{j,i}) \over 2{d(i) \over vol(G')} {d(j) \over vol(G')}} \nonumber \\
&={vol(G') \over 2} \sum_{r\in \{1,3,...,2n-1\}} \bigg(  {1 \over d(j)} \cdot (\mathbf{P^{r}})_{i,j} + {1 \over d(i)}\cdot (\mathbf{P^{r}})_{j,i}\bigg) \end{align}
In matrix form the right hand side of Equation \eqref{eq:intp} is equivalent to $(i,j)th$ entry of the following matrix
\begin{align}
\label{eq:finalmatrix}
&{vol(G') \over 2} \sum_{r\in \{1,3,...,2n-1\}} \bigg(  \mathbf{P^{r}} D^{-1} + D^{-1}(\mathbf{P^{r}})^{T}\bigg) \nonumber \\
=& {vol(G') \over 2} \sum_{r\in \{1,3,...,2n-1\}}(  (D^{-1}\mathcal{A})^r D^{-1} + D^{-1}(\mathcal{A} D^{-1})^r ) \nonumber \\
=& {vol(G')} \sum_{r\in \{1,3,...,2n-1\}} (D^{-1}\mathcal{A})^r D^{-1}.
\end{align}
\end{proof}
We emphasize that unlike equivalence proofs in previous works \cite{qiu2017network}, we do not make any assumptions about infinite long walks and stationary distributions over undirected and non-bipartite graphs because of the following facts. Firstly, the initial distribution for the first vertex $v$ is $d^{out}(v)\over vol(G)$ for a source walk and $d^{in}(v)\over vol(G)$ for a target walk, unlike the uniform distribution used in other methods. Secondly, we use the first vertex of the walk as the input vertex and we know the exact distribution from which it is drawn. As a result, the distribution for training pairs can be computed analytically. 

\subsection{Connection with SALSA and HITS}

As already mentioned, \nerd derives its motivation from the classical HITS and SALSA algorithms which are based on the idea that in all types of directed networks, there are two types of important nodes: {\em hubs} and {\em authorities}.
Good hubs are those which point to many good authorities and good authorities are those pointed to by many good hubs. We also base our \nerd model on a similar intuition, in which we aim to embed nodes co-occurring in alternating walks closer in their respective source and target vector spaces. Moreover, each iterative step of HITS requires hub/authority scores to be updated based on authority/hub scores of neighboring nodes. NERD attempts to extend this by exploring slightly bigger neighborhoods by embedding source target pairs closer if the co-occur in walks of some small size $\ell$.

Technically, given a directed graph $G$ and adjacency matrix $A$, HITS is an iterative power method to compute the dominant eigenvector for $A^TA$ and for $AA^T$.  
The authority scores are determined by the entries of the dominant eigenvector of the matrix $A^TA$, which is called the {\em authority matrix} and the hub scores are determined by the entries of the
dominant eigenvector of $AA^T$, called the {\em hub matrix}.  This is equivalent to finding the dominant eigenvector of the matrix $\mathcal{A}^2$ (where $\mathcal{A}$ is constructed as in Equation (1) in the paper) , the first $N$ entries then correspond to hub scores and the later to authority scores. SALSA instead considers the transition matrix given by $D^{-1}\mathcal{A}$ and corresponds to computing hub and authority scores using the principal eigenvector of $(D^{-1}\mathcal{A})^2$.
\subsection{Differences with the Previous Approaches}

Though various ingredients of \nerd already existed, it combines them in a novel way. We note that the concept of alternating walks already dates back to classical algorithms like HITS and SALSA but its use to preserve role information for node embeddings and sample neighborhoods to generate low dimensional representations is the first of its kind. 

Again there are subtle but important differences in modeling of our objective function and its optimizing using negative sampling. First, we use source and target alternating walks as against directed walks only following the outgoing links in all previous approaches. Note that using only  walks in one direction will never have walks sampled from vertices with $0$ outdegree. Additionally if these vertices also have indegree, these might be sampled in no or a few training examples, hence leading to bad quality of their embeddings. 

Second, we recall that most of the previous methods learn node and context matrices (in a similar fashion as word2vec) but uses only the node matrix for downstream tasks. The node matrix is the matrix from input to hidden layer and the context matrix is the one from hidden to output layer. \app which also uses context matrix for downstream tasks follows the same architecture, wherein the context matrix is used to represent the embedding vectors of a vertex in its destination role. In principle it uses the vertex in its source role as input and tries to predict the sampled neighbor which is always considered to be in its destination role. We on the other hand use the architecture in both directions using the output/input layer as input/output when our input/output vertex is in its destination/source role.
For example after the target walk, the input is the vertex in the destination role (causing updates to context matrix) and the predicted vertex is in its source role (hence causing updates to vertex matrix). 

\section{Parameter Settings}
For fair comparisons, the embedding dimensions were set to $128$ for all approaches.For node classification, embeddings of $64$ dimensions each were concatenated for methods using two embedding matrices. \lineemb-1+2 corresponds to first normalization of 64 dimension embedding matrices of \lineemb1 and \lineemb2 followed by concatenation (as suggested in the paper).

For \hoppe, the attenuation factor $\beta$ was set to $0.01$ across all datasets and tasks except PubMed for which $0.5$ was used. 

For \nodetovec, we run experiments with walk length \(l=40\), number of walks per node \(r = 80\), and window size \(10\), as described in the paper. The results are reported for the best in-out and return hyperparameters selected from the range $p,q \in \{ 0.25,0.5,1,2,4\}$.
In particular, the reported results correspond to the following in-out and return parameters: $p=4, q=4$ for link prediction, $p=0.25, q=4$ for multi-label classification and  $p=0.25, q=4$  for graph reconstruction tasks in the Cora dataset; $p=0.25, q=1$ for link-prediction and graph reconstruction tasks in Twitter.

For \verseemb, we use $\alpha=0.85$ for all tasks across all datasets as suggested in the paper. For \deepwalk,the parameters described in the paper are used for the experiments, i.e.\ walk length \(t=40\), number of walks per node \(\gamma = 80\), and window size \(w=10\). For \lineemb, we run experiments with total \(T=10\) billion samples and \(s=5\) negative samples, as described in the paper.  For the multi-label classification task, the two halves of dimensions 64 (the embeddings from \lineemb1 and \lineemb2) are normalized and concatenated. For \app we used the restart probability as $0.15$ and $80$ as number of walks per vertex (as also used in other works) and $10$ as the number of samples per vertex, giving total number of walks per vertex as $800$. No exact numbers for these parameters were provided in the original paper. For fair comparisons we also fixed the total number of walks in \nerd as $800$ times the number of vertices. 
For \nerd, the mini-batch size of stochastic gradient descent is
set to $1$ walk sample, i.e.\ $n$ input-neighbor pairs. The learning rate
is set with the starting value $\rho_0 = 0.025$ and $\rho_t = \rho_0 \left( 1-t/T \right)$ where $T$ is the total number of walk samples. The number of walk samples is fixed to $800$ times the number of vertices. Other parameters, i.e., number of neighborhood nodes to be sampled $n$ and number of negative samples $\kappa$, could in principle vary over datasets and across tasks and can be fine tuned using a small amount of training data. For link prediction $n=1,\kappa=3$ was used across all datasets, for graph reconstruction and node classification $n=10,\kappa=5$ was used along with joint training of nodes of similar roles.

\section{Supplementary Experiments} 
\label{sec:supplementary}
\subsection{Learning Asymmetrical Edge Representations} The approach presented in \cite{learnEdge2017} uses a deep neural network (DNN) to obtain edge representations from trainable node embeddings as inputs. This method also uses a simple embedding space for representing nodes. Specifically, the DNN learns to output representations that maximize the \emph{Graph Likelihood}, which is defined as the overall probability of correctly estimating the presence or absence of edges in the original graph, using (trainable) node embeddings as inputs. \begin{table}[h!] \begin{center} \footnotesize \setlength{\tabcolsep}{1pt} \renewcommand{\aboverulesep}{0pt} \renewcommand{\belowrulesep}{0pt} \newcolumntype{C}{>{\centering\arraybackslash}X} \begin{tabularx}{\columnwidth}{XCCCC} \multicolumn{1}{c}{} & \multicolumn{4}{c}{\textit{Training Data, \% Edges}}\\ \cmidrule{2-5} \multicolumn{1}{l}{\emph{method}} &20\% & 50\% & 70\% & 90\% \\ \midrule \small\edgednn & 0.719 & 0.722 & 0.716 & \textbf{0.753}\\ \bottomrule \end{tabularx} \end{center} \caption{Link prediction results (AUC scores) in the CORA dataset.}\label{tab:link-pred-edge-dnn} \end{table} We run the authors' implementation of the approach on the CORA dataset using multiple train-test-splits created by their method ( to provide them the advantage). The AUC scores resulting from link prediction evaluation are presented in Table \ref{tab:link-pred-edge-dnn}. The training for Twitter dataset did not finish after running for $1$ day. The results show that the method is performing worse than \nerd. Moreover, it uses a much more complex architecture than \nerd which is difficult to fine-tune for a variety of tasks. For example no leverage over other undirected methods can be achieved in node classification task as this method encode asymmetry information of edges and not the nodes.

\subsection{Parameter Sensitivity}
\begin{figure*}[t]
\centering
    \subfloat[Dimensions Vs. Micro-F1]{\includegraphics[width=0.3\textwidth]{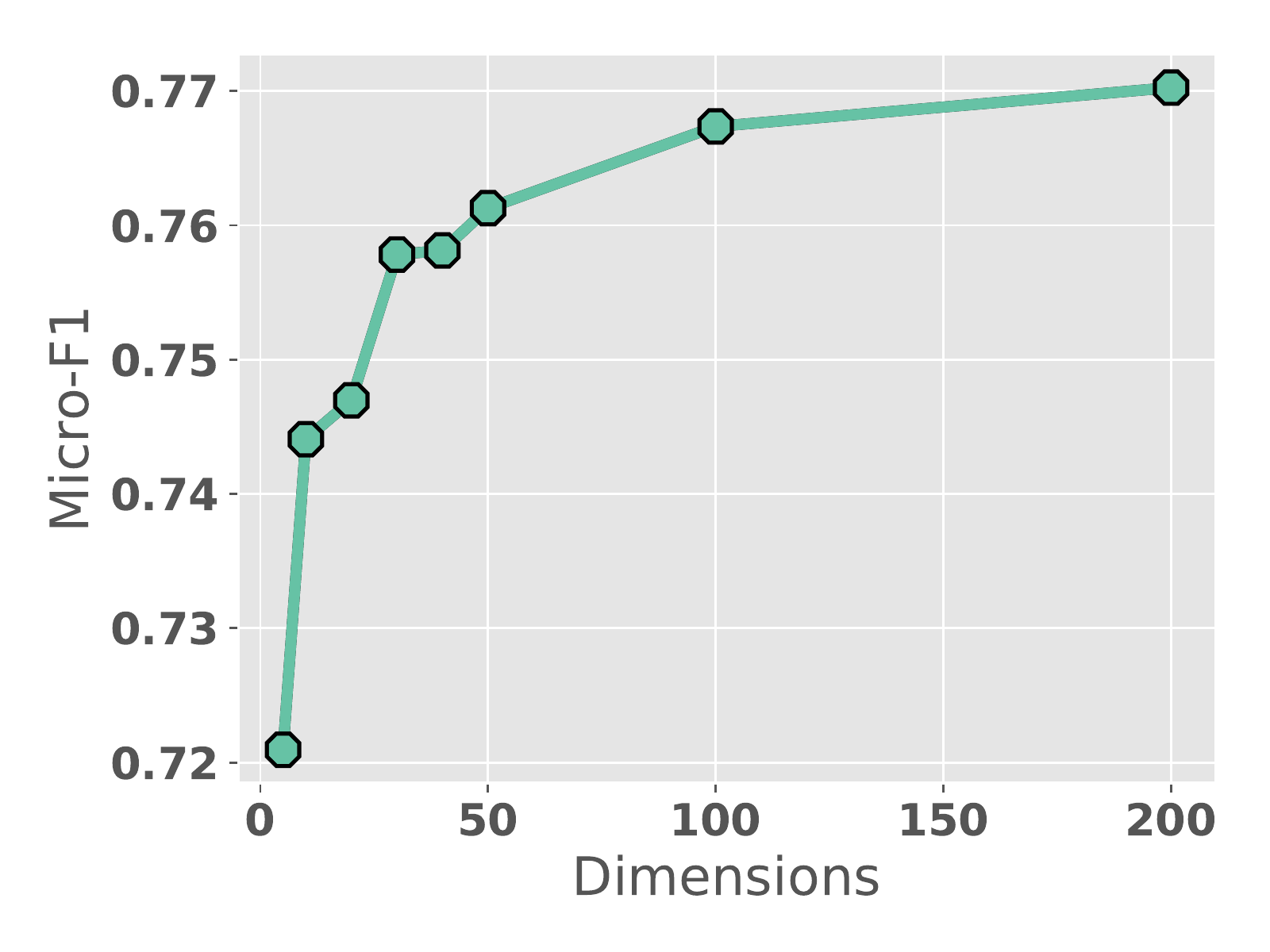}}
    \subfloat[Number of samples (in Million) Vs Micro-F1]{\includegraphics[width=0.3\textwidth]{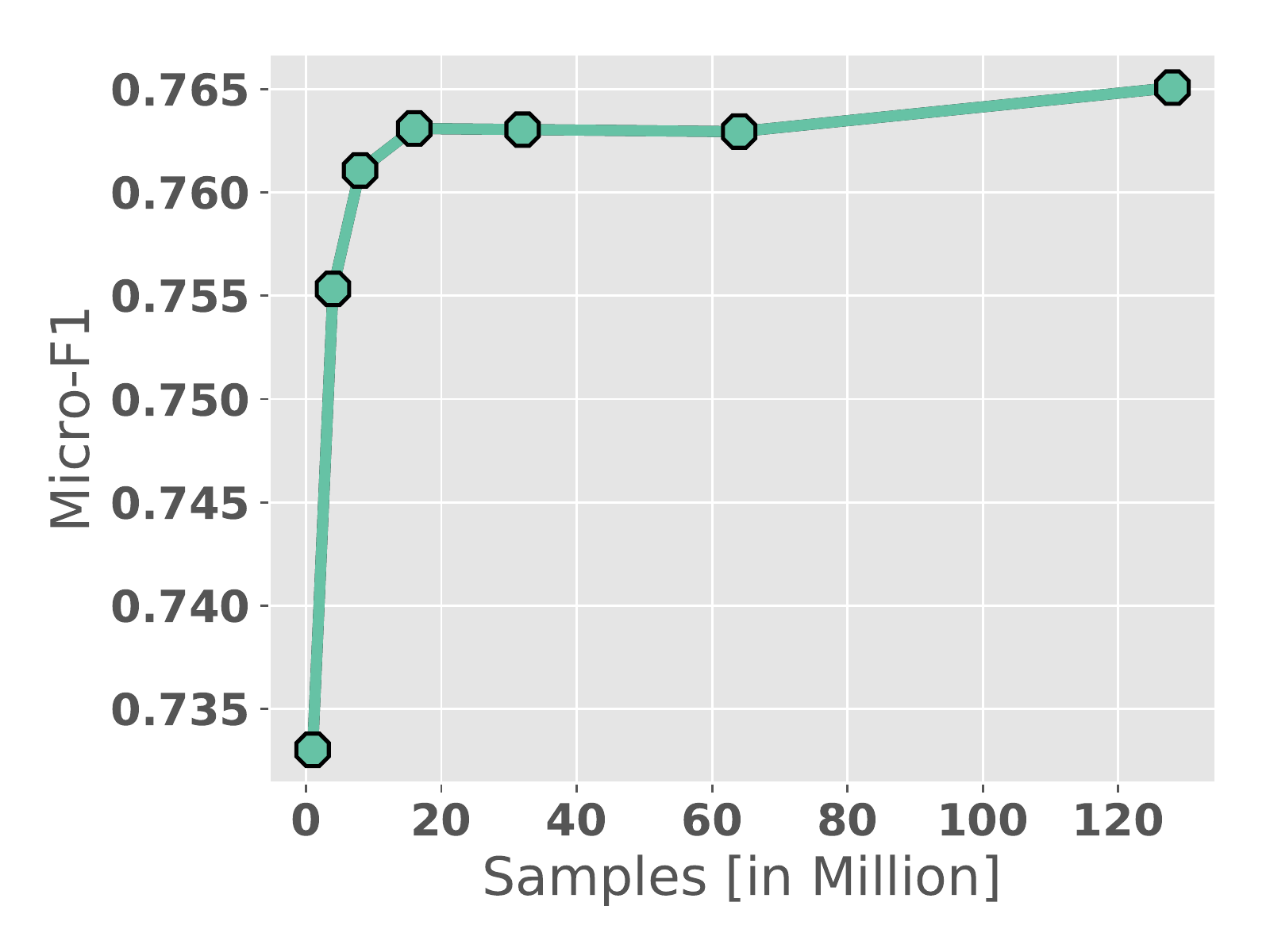}}
\caption{Parameter Sensitivity Plots. Effect of changing hyperparameters on Node classification results in PubMed}
\label{fig:parameterSensitivity}
\end{figure*}

 We next investigate the performance of \nerd with respect to the embedding dimensions and its converging performance with respect to the number of walk samples on the multi-label classification task in the PubMed dataset (see Figure \ref{fig:parameterSensitivity}).
 We note that \nerd achieves a good performance better than \deepwalk at 30 dimensions (as embeddings are concatenated, the total dimensions are 60).
The performance of \nerd converges quite fast with the number of walk samples. Note that the reported performance for PubMed in Table 3 corresponds to 16M samples.

\begin{figure}[t]
\centering
    {\includegraphics[width=0.45\textwidth]{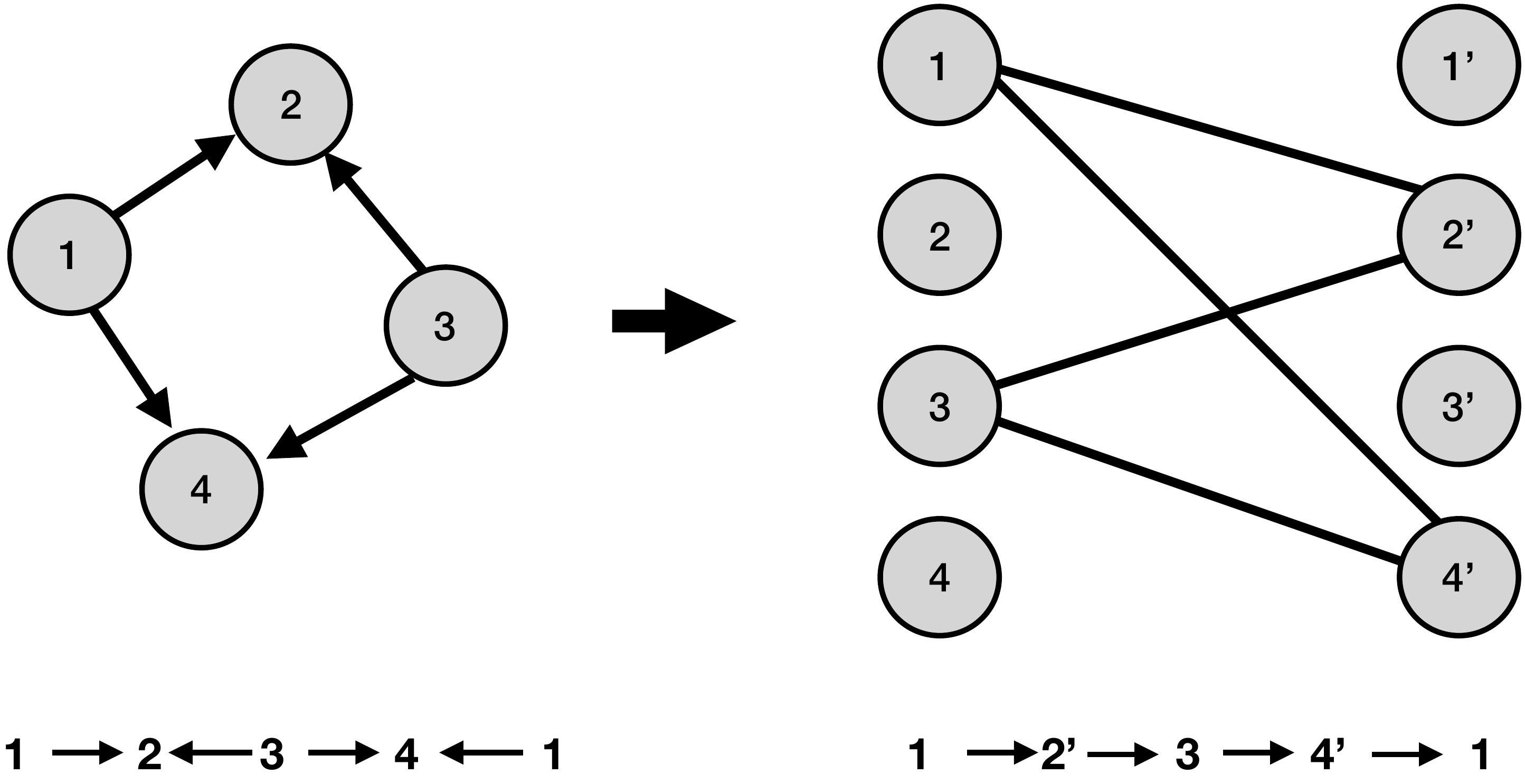}}
\caption{Alternating walk on a directed graph (left) can be seen as a walk on an equivalent bipartite graph (right).}
\label{fig:bipartite}
\end{figure} 
\end{appendix}
\end{document}